\documentclass[10pt,journal,compsoc]{IEEEtran}

\usepackage{times}
\usepackage{soul}
\usepackage{url}
\usepackage[hidelinks]{hyperref}
\usepackage[utf8]{inputenc}
\usepackage[small]{caption}
\usepackage{graphicx}
\usepackage{amsmath}
\usepackage{amsthm}
\usepackage{booktabs}
\usepackage{algorithm}
\usepackage{algorithmic}
\usepackage[switch]{lineno}

\usepackage{amsfonts}
\usepackage{amssymb}
\usepackage{url}
\usepackage{booktabs}
\usepackage{graphicx}
\graphicspath{{Figures/}}
\usepackage{bm}
\usepackage{pbox}
\usepackage{mathrsfs}
\usepackage{enumitem}
\usepackage{subfigure}
\usepackage{color, colortbl}
\usepackage{comment}
\usepackage{mathtools, cuted}
\usepackage{bbm}
\usepackage{dsfont}

\newtheorem{theorem}{Theorem}
\newtheorem{corollary}{Corollary}
\newtheorem{lemma}{Lemma}
\newtheorem{remark}{Remark}

\newtheorem{assumption}{Assumption}

\author{Ying~Cao,
        Siyuan~Yu,
        Xiaoqi~Tan,
        and~Danny~H.K.~Tsang
\IEEEcompsocitemizethanks{
\IEEEcompsocthanksitem Y. Cao is with the Department of Electronic and Computer Engineering, HKUST, HKSAR.\protect\\
E-mail: ycaoan@connect.ust.hk

\IEEEcompsocthanksitem{Siyuan~Yu and~Xiaoqi~Tan are with the Department of Computing Science, University of Alberta, Edmonton, AB, Canada.\protect\\
E-mail: syu3,~xiaoqi.tan@ualberta.ca}

\IEEEcompsocthanksitem{Danny H.K. Tsang is with the Thrust of Internet of Things, HKUST(GZ), GZ, China.\protect\\
E-mail: eetsang@hkust-gz.edu.cn}%

}}

\begin{document}

\title{Competitive Analysis of Online Path Selection: Impacts of Path Length, Topology, and System-Level Costs}

\IEEEtitleabstractindextext{

\begin{abstract}
Consider a communication network to which a sequence of self-interested users come and send requests for data transmission between nodes. This work studies the question of how to guide the path selection choices made by those online-arriving users and maximize the social welfare. Competitive analysis is the main technical tool. Specifically, the impacts of path length bounds and topology on the competitive ratio of the designed algorithm are analyzed theoretically and explored experimentally. We observe intricate and interesting relationships between the empirical performance and the studied network parameters, which shed some light on how to design the network. We also investigate the influence of system-level costs on the optimal algorithm design.
\end{abstract}

}

\maketitle

\IEEEdisplaynontitleabstractindextext
\IEEEpeerreviewmaketitle
\IEEEraisesectionheading{\section{Introduction}\label{sec:introduction}}


\IEEEPARstart{F}{rom} communication networks to transportation networks, strategically allocating network resources to competing users in a decentralized way is a recurring theme in network research. Network operators design resource allocation strategies to achieve specific goals, including revenue maximization, social welfare maximization, and cost minimization. In transportation networks, designing efficient local routing strategies for vehicles to best accommodate to users' dynamic demands is one of the key challenges. In parallel, in communication networks, routing, congestion control and scheduling are practical real-time resource allocation mechanisms to improve the network performance.
This work focuses on routing data traffic in communication networks. Following the convention of economics, data source nodes are considered as self-interested agents who demonstrate sufficient autonomy and pure rationality when making decisions. Each agent decides to send data through the network or not and select routing paths at will, as is the case with source routing in the future Internet~\cite{wg2018source}. Specifically, we are concerned about how to navigate data traffic from different agents through the network such that the social welfare is maximized. 


One concern among others that network operators typically have is the allocation efficiency of the limited network capacity. Given the welfare of each agent for routing her data through a certain path, integer linear programming can model and deal with this concern. However, the set of agents in the network is usually dynamic. Which agents will join is generally unknown to the network operator \textit{a priori} -- multiple methodologies are devoted to dealing with this uncertainty. Each methodology makes a different set of assumptions. Assuming data -- e.g., the welfare and resource consumption level of each agent -- follow an either known or unknown probability distribution, one can employ stochastic optimization methods to provide a performance guarantee if an underlying distribution exists but little can be guaranteed when outliers don’t belong to the modelled distribution. Thus, to avoid the dependency on any stochastic assumption, we follow the worst-case analysis framework, which is a robust modeling and analysis framework that makes the least assumptions about the environment.

The worst-case analysis framework compares the performance of a decision-maker who is uninformed about the future with the performance of a hypothetical oracle who has perfect foresight and can make optimal decisions. It provides a comparative performance guarantee that holds even in the worst-possible scenario. In other words, this framework assumes an adversary who can manipulate the sequential data and the environment. Two performance measures dominate this field of analysis -- the competitive ratio and the adversarial regret. The primary difference between them is the type of performance guarantee they provide. The regret provides an additive guarantee, bounding the difference between the decision-maker's performance and the optimal one. In contrast, the competitive ratio offers a multiplicative guarantee, meaning the decision-maker's performance is bounded by a constant factor of the optimal performance.
This work adopts the competitive ratio as the performance measure of interest.

Another concern of network operators is the performance evaluation of routing algorithms under various network configurations~\cite{yang1997performance}. By understanding the impacts of network configurations on the algorithmic performance, one can better configure the network to improve. For instance, the more nodes connecting the source and destination nodes, the longer the time it takes to transfer the data. Taking time as a limited and precious resource, network operators usually prefer the shortest-path routing which minimizes the data transfer time. Also, the average path length of a network topology, defined as the average path length between any two nodes, is usually linked to the easiness for communication within the network~\cite{ye2010distance}. For example, a larger average path length in a communication network indicates a slower and less efficient information transfer process. Thus, to investigate the impact of the path length and the topology on the social welfare, we assume that possible path lengths fall into a known interval, characterizing the uncertainty perceived by the network operator in users’ path lengths, and ask the question, \textit{how does the uncertainty in path lengths affect the social welfare in different topological networks?}

In addition to network configuration parameters, the social welfare is also deeply influenced by the presence of costs. For example, the queuing delay experienced by users is usually viewed as a type of cost. Packets will be discarded by applications if their waiting times are unacceptably long, and the user experience is usually inversely related to the experienced delay. An innate property of the queuing delay is that it is experienced and influenced by all agents in the system, leading to coupling effects between agents. A well-designed admission control or packet scheduling strategy can exert a distributed system-level control over the queuing delay and improve the overall service experience. With more agents joining the system, the social welfare is increased from serving more demands, while the competition becomes tenser, and thus a higher cost, decreasing the social welfare. Queuing theory has been partially devoted to understanding and characterizing the queuing delay within the system. We incorporate the mathematically well-defined queuing delay therein as a cost and investigate the influence of such system-level costs on the algorithm design. 

\subsection{Related Work}
\textbf{Online Competitive Resource Allocation.} The worst-case analysis technique of this paper falls under the umbrella of the competitive analysis framework~\cite{borodin2005online}. There exists a series of works closely related to the problem studied here, such as competitive online routing~\cite{bose2004competitive,buchbinder2006improved} and variants of online knapsack~\cite{sun2022online,yang2021competitive}. Surprisingly, none of the above works studies either the impact of the topology or the path length on the performance, which partially motivates this work. In comparison, online resource allocation with costs has been studied under different scenarios. For example, online combinatorial auction with convex costs was studied in~\cite{tan2020mechanism}, and welfare maximization with polynomial costs was studied in~\cite{huang2019welfare}. However, a cost function that goes to infinity when reaching the capacity limit, such as the queuing delay, has not been studied before.

\textbf{Routing Games.} Similar routing problems have also been studied via the lens of game theory, such as selfish routing games~\cite{roughgarden2002bad,banner2007bottleneck} and network congestion games~\cite{hao2024inefficiency}. The performance measure of interest is the so-called price of anarchy, which compares the system performance where agents make decisions locally and selfishly with that produced by a centralized optimizer. It characterises the inefficiency of an equilibrium caused by the absence of a centralized regulator. In general, prior works mentioned here are concerned with the decentralized aspect of the problem, specifically, the interplay between agents, while this work is focused on retaining competitive against the uncertainty of the future. It is worth mentioning that the combination of these two aspects is an interesting and challenging direction where more attention is deserved.

\textbf{Regret Analysis.}
The regret offers an alternative additive guarantee to the decision maker against the uncertainty of future. The field of online convex optimization dedicates to analyzing the regret compared with static or dynamic benchmarks. The online convex optimization with constraints is close to our system setting. In~\cite{chen2017online}, they studied an online convex optimization problem with time-varying constraints and compared the designed online algorithm with a dynamic benchmark. Regret and constraint violations were analyzed for the proposed algorithm. The constraints in our setting are time-varying but \textit{coupled across time slots}, and we do not allow constraint violations. Moreover, it has been explored that best-of-both-worlds guarantees, i.e., a sublinear regret and a bounded competitive ratio at the same time, can be achieved under specific settings, such as in metrical task systems~\cite{daniely2019competitive}. To the best of our knowledge, it remains unknown whether best-of-both-worlds guarantees exist in other online problems.


\textbf{Online Algorithm with Constrained Adversary.} Another recent trend in the field of online algorithms is to consider a constrained adversary. Instead of granting the adversary arbitrary power to manipulate future arrivals, additional control could be exert on the possible arrival instance. For example, future prices were restricted in a known interval in the online selection problems~\cite{jiang2021online}. Under different constrained adversaries, the design strategy of competitive online algorithms could be completely different. For example, in the online selection problem, an adversary that constrains the price range and one that constrains the horizon admits completely different competitive algorithms. This work can be viewed as designing competitive algorithm against the adversary who is constrained on the path length and the topology.


\subsection{Contributions and Paper Organization}
Our main contribution in this work is as follows. First, we consider the online path selection problem with constraints on path lengths and network topology. A set of topology-dependent competitive ratios are derived for the considered line and tree topologies. Second, we study the impacts of the minimum path length $ m $ for the first time and recover existing results about the logarithmic dependence of the competitive ratio on the maximum path length $ M $. We show that the minimum path length $ m $ plays a role of varying importance in different networks. Specifically, the competitive ratio decreases slower with $m$ in tree networks compared to line networks, which also highlights the influence of the topological structure. Finally, we conduct extensive experiments to examine our theoretical findings and explore the performance of different price aggressiveness, which is a parameter in the designed algorithm. Our results show that the relationship between the empirical ratio and the path length bound varies for different topologies and different price aggressiveness, for which we provide an in-depth analysis and offer insights into the network design.

The rest of this paper is organized as follows. Section 2 introduces the system model and derives competitive ratios that are dependent on the network topology and the path length. Specifically, two fundamental networks, line networks and hierarchical tree networks, are studied. We also provide an analysis on the impact of system-level costs on the competitive algorithm design in this section. In Section 3, we conduct extensive experiments to examine the empirical performance of the online algorithm. Specifically, we first examine the gap between worst-case theoretical guarantees and empirical performance against stochastic arrivals, and then we verify the theoretical results by running the algorithm on certain hard instances. For the path selection with cost problem, we show the logarithmic trend of the competitive ratio with respect to the maximum value density by numerical methods. 
Finally, we conclude the paper in Section 4.

\section{System Description and Results}\label{sec:system}
The system of concern is described as follows. We consider a network $ \mathcal{N} $ whose topology is fixed. Nodes in the network can be viewed as routers or servers. 
Edges connecting nodes are endowed with fixed capacities, which refers to capacities of fixed electronic wires. 

(\textbf{Arrival Instance})
Agents who need to be routed from one node to another node arrive at the network in an online fashion. There are $N$ agents in total, but the decision-maker does not know the value of $N$. An agent submits a reservation request on her arrival to use a certain bandwidth for transmitting data along some routing path. The strategic setting is considered here, where each agent holds a private value of her demand if fulfilled and could report false values to the network for her benefit. In summary, each agent can be characterized by the following parameters:
\begin{itemize}
    \item Value: $ v_i $
    \item Rate requirement: $ r_i $
    \item Source and destination nodes: $s_i$ and $t_i$
    \item A set of possible paths connecting $s_i$ and $t_i$: $ \mathbb{P}_i $
\end{itemize} 
Without the loss of clarity, we use agents and requests interchangeably throughout the paper.

\begin{assumption}[Small Request Size]\label{assumpt: max-size}
    The rate requirement of any agent is upper-bounded by $ \epsilon \in \mathbb{R}^+$.
\end{assumption}

Assumption~\ref{assumpt: max-size} is self-explanatory in many applications. For example, in cloud data centers where individual workloads are negligible compared to the computing capacity of servers. In communication networks with a considerable size, it is also expected that individual requests is negligible in comparison to the transmission capacity of any edge in the network. It is also important for the theoretical analysis, for example, in the proof of Theorem~\ref{thm: upbds-line}.

\begin{assumption}[Bounded Path Length]\label{assump: pathlength}
    The path that a request can select is bounded in its length, i.e., $ m \le |P_i| \le M $.
\end{assumption}
Assumption~\ref{assump: pathlength} further constrains the set of possible arrivals by limiting the number of edges on any routing path.

In addition, any reasonable agent should not hold an infinite value of a finite rate. The following modeling assumption is to formalize this observation.
\begin{assumption}[Bounded Value Density]\label{assumpt: net-cent}
    For any agent $i$, the ratio between her value and her \textbf{total} resource consumption is bounded, i.e., $ \frac{v_i}{|P_i| r_i} \in [1,\bar{p}]$.
\end{assumption}

Assumption \ref{assumpt: net-cent} sets uniform bounds on the per-unit value of any request, which leads to the value of user $i$ in proportion to the number of edges on path $ P_i $, i.e., $ v_i \in [d_i, d_i \bar{p}]$, where $ d_i $ is the path length of $P_i$. This is a well-accepted assumption for various online decision problems such as the online time series search~\cite{el2001optimal,tan2023threshold} and online knapsack problem~\cite{sun2022online,yang2021competitive}, etc. Online path selection can be viewed as a special case of online knapsack problem~\cite{cao2024competitive}.


Following the standard competitive analysis framework, the performance of an online algorithm is characterized by the \textit{competitive ratio}, defined as
\begin{align*}
    \max_{I\in\Omega} \frac{\textsf{OPT}(I)}{\textsf{ALG}(I)},
\end{align*}
where $I$ denotes an instance and $ \Omega $ represents the family of instances that satisfy Assumptions \ref{assumpt: max-size}-\ref{assumpt: net-cent}.

\subsection{Posted-Price Mechanism}
We propose Algorithm~\ref{alg:generic-path-select} below, whose nature is a posted-price mechanism. A central operator maintains a price for each edge in correspondence to its utilization level and announces the edge prices publicly. When an agent arrives, she calculates the minimum price over all paths that can fulfill her demand, compares it with her private value, and decides to join or leave the network based on the comparison. Importantly, a posted-price mechanism is both incentive-compatible and privacy-preserving as the agent does not need to report her private value to a central operator~\cite{feldman2014combinatorial}.
\begin{algorithm} 
\caption{Posted-Price Mechanism for Path Selection (\textsf{PPM-PS}$_{\phi}$)} 
\label{alg:generic-path-select} 
\begin{algorithmic}[1] 
    \REQUIRE Utilization $\omega_e^{(i)} \leftarrow 0$.
    \WHILE{A new agent $i$ arrives}
        \STATE {Find the path with the minimum price:
        $$ P_i = \arg\min_{P\in \mathbb{P}_i} r_i \lambda^{(i-1)}(P). $$}
        \IF{ $ v_i > r_i \lambda^{(i-1)}(P_i)$}
            \IF{$ \omega_e^{(i-1)}+r_i \le C_e, \forall e \in P_i $ }
            \STATE {Join the network in full on $P_i$}
            \FOR{each edge $e$ on $P_i$}
                \STATE {$ \omega_e^{(i)} = \omega_e^{(i-1)} + r_i $. }
                \STATE {Update edge price $ \lambda^{(i)}_e = \phi_e \left(\omega_e^{(i)}\right) $.}
            \ENDFOR
            \ENDIF
            \STATE {Leave the network.}
        \ENDIF
    \ENDWHILE
\end{algorithmic}
\end{algorithm}

The price of a path $P$, i.e., $\lambda^{(i-1)}(P)$, is calculated by summing over the price of all edges on this path, i.e., $\lambda^{(i-1)}(P) = \sum_{e\in P}\lambda^{(i-1)}_e$, and the pricing function
$ \phi_e(\cdot) $ is parameterized by a parameter $\gamma$: $ \phi_e^{\gamma}(\omega) = e^{\gamma\omega/C_e} - 1 $.
The exponential pricing function is a classic choice in online algorithm design, where $\gamma$ is a parameter that guides the aggressiveness of the price. If $\gamma$ is larger, the price increases much faster with the utilization, indicating that more capacity is reserved for the future, and thus leading to a more conservative allocation.


\subsection{Main Results I: Topology-Dependent Theoretical Guarantees of \textsf{PPM-PS}$_{\phi}$
}
To understand what effect the network structure imposes on the performance of the mechanism \textsf{PPM-PS}$_{\phi}$, a set of fundamental and amiable network topologies, i.e., line networks and tree networks, are studied in this subsection. 

\subsubsection{Line Network}
Line networks are the simplest networks.
We consider a bi-directional line network with $ N+1 $ nodes and $ N $ edges. The following theorem shows the competitive ratio of \textsf{PPM-PS}$_{\phi}$ for a line network.

\begin{theorem} \label{thm: upbds-line}
For line networks,  when $\epsilon \le \frac{C_{\min}}{\gamma}$, \textsf{PPM-PS}$_{\phi}$ is $ \max\{O(\ln M\bar{p}), O(\beta \ln (\frac{M\bar{p}}{2m \beta}+1))\} $-competitive, where $ \beta $ is the ratio between the largest capacity and the smallest capacity in the network.
\end{theorem}

\begin{proof}
    
For line networks, partition the line network $ \mathcal{N} $ into $ J = \lfloor \frac{N}{M} \rfloor $ disjoint lines $ \mathcal{N}=\cup_j \mathcal{N}_j $, each containing $M$ consecutive edges and $M+1$ nodes. Requests can travel through at most two $M$-long lines. Those starting from $\mathcal{N}_j$ and ending in $ \mathcal{N}_{j+1} $ are grouped in $I_{j}^1$, and those starting from $\mathcal{N}_j$ and ending in $ \mathcal{N}_{j-1} $ are grouped in $I_{j}^2$. Let $I^1 = \cup_j I_{j}^1$, $I^2 = \cup_j I_{j}^2$ and $ I = I^1 \cup I^2$. Because the data transmission is usually unidirectional, requests in $I^1$ do not affect those in $I^2$. Built on this decoupling, we have 
\begin{align*}
    \frac{\textsf{OPT}(I)}{\textsf{ALG}(I)} &= \frac{\textsf{OPT}(I^1) + \textsf{OPT}(I^2)}{\textsf{ALG}(I^1) + \textsf{ALG}(I^2)} \\
    &\le \frac{\textsf{OPT}(I^1)}{\textsf{ALG}(I^1)} + \frac{\textsf{OPT}(I^2)}{\textsf{ALG}(I^2)},
\end{align*}
where $ \textsf{OPT}(I) $ and $ \textsf{ALG}(I) $ denotes the optimal revenue and the revenue of the online algorithm given the instance $I$. To simplify the notation, we drop the superscripts $1$ and $2$ from now on and focus on requests in one direction, e.g., $I_{j}$ is the set of requests that start from line network $\mathcal{N}_j$.

Let $ \tilde{I}_{j} = I_{j-1} \cup I_{j}, j\ge 2 $. All requests in $ \tilde{I}_{j} $ affect edges in $\mathcal{N}_{j} $. In the sequel, we focus on the upper bound for the ratio $ \frac{\textsf{OPT}(I_{j})}{\textsf{ALG}(\tilde{I}_{j})} $ over instances in any $M$-long line network $ \mathcal{N}_{j} $ because the competitive ratio is upper-bounded as:
\begin{align*}
    \frac{\textsf{OPT}(I)}{\textsf{ALG}(I)} &= \frac{\sum_{j} \textsf{OPT}(I_{j})}{\sum_{j} \textsf{ALG}(I_{j})} \\
    &= \frac{2 \sum_{j} \textsf{OPT}(I_{j})}{\textsf{ALG}(I_{1}) + \sum_{j} \textsf{ALG}(\tilde{I}_{j}) + \textsf{ALG}(I_{J})} \\
    &\le \max_{j} \frac{2 \textsf{OPT}(I_{j})}{\textsf{ALG}(\tilde{I}_{j})},
\end{align*}
Let $\tilde{\mathcal{N}}_j = \mathcal{N}_{j-1} \cup \mathcal{N}_{j} \cup \mathcal{N}_{j+1} $. Any edge in $\tilde{\mathcal{N}}_j$ can be possibly affected by requests in $ \tilde{I}_{j} $. When there is no saturated edge in $ \tilde{\mathcal{N}}_j $, i.e., no edge $ e \in \tilde{\mathcal{N}}_j $ with $ \omega_e^{(i)} > C_e - \epsilon $,
after the $i$th request with $P_i \subset \tilde{\mathcal{N}}_j$ is routed, the value generated by the online algorithm increases by $\Delta \textsf{ALG} = v_i$. The increase to the optimal solution $ \Delta \textsf{OPT} $ is upper-bounded by $\Delta C$ based on the weak duality, and we have
\begin{align*}
    \Delta C &= \sum_{e\in P_i} C_e \left(\lambda_e^{(i)} - \lambda_e^{(i-1)} \right) + v_i \\
    &= \sum_{e\in P_i} C_e \left(\lambda_e^{(i-1)}+1 \right) \left[\exp \left(\frac{r_i \gamma}{C_e} \right)-1 \right] + v_i \\
    &= \sum_{e\in P_i} r_i\left(\lambda_e^{(i-1)}+1 \right)\frac{C_e}{r_i} \left[\exp \left(\frac{r_i \gamma}{C_e} \right)-1 \right] \\
    &+ \Delta \textsf{ALG}.
\end{align*}
It follows that 
\begin{align*}
    \Delta C &\le \frac{C_{\min}}{r_i}\left[ \exp \left(\frac{r_i \gamma}{C_{\min}} \right) - 1 \right] \sum_{e\in P_i} r_i \left(\lambda_e^{(i-1)}+ 1 \right)\\
    &+ \Delta \textsf{ALG} \\
    &\le \frac{C_{\min}}{r_i}\left[ \exp \left(\frac{r_i \gamma}{C_{\min}} \right) - 1 \right] (v_i+r_i d_i) + \Delta \textsf{ALG} \\
    &\le \left\{ 2\frac{C_{\min}}{r_i}\left[ \exp({\frac{r_i \gamma}{C_{\min}}}) - 1 \right] + 1 \right\} \Delta \textsf{ALG}
\end{align*}
because $r_i \sum_{e\in P_i} \lambda_e^{(i-1)} \le v_i$, $\frac{v_i}{r_i d_i}\in [1,\bar{p}]$, and $x \left[ \exp(\frac{\gamma}{x})-1\right ]$ is decreasing in $[\gamma,\infty)$. When $r_i \le  \epsilon \le \frac{C_{\min}}{\gamma} $, we have $\frac{\Delta C}{\Delta \textsf{ALG}} \le 2\frac{C_{\min}}{r_i}\left( e^{\frac{r_i \gamma}{C_{\min}}} - 1 \right) + 1 \le 2(e-1)\gamma + 1 $. Summing over $i$, we have $\textsf{OPT}(I_{j})\le C(I_{j})$, and $ \frac{\textsf{OPT}(I_{j})}{\textsf{ALG}(\tilde{I}_{j})} \le \frac{C(I_j)}{\textsf{ALG}(\tilde{I}_{j})} = 2(e-1)\gamma + 1 $. Therefore, the competitive ratio for the case without edge saturation is $ 4(e-1)\gamma + 2 $.

When there exist almost-saturated edges (case with edge saturation), i.e., $\exists \tilde{e} \in \tilde{\mathcal{N}}_j, w_{\tilde{e}}^{(i)} > C_{\tilde{e}}-\epsilon$, we need to show that the algorithm output is still lower-bounded by a portion of the optimal value, when requests are rejected due to the capacity limit instead of an insufficient value. Corollary~\ref{corol: alg-lb-opt-expr} characterizes the largest possible lower bound by an optimization problem. Then we use Corollary~\ref{corol: cr-overall} to bound the competitive ratio. Both corollaries are useful for the analysis of the tree network as well.

\begin{corollary}\label{corol: alg-lb-opt-expr}
    If there exists a link $\tilde{e}$ whose $\lambda_{\tilde{e}}^{(N)}$ is close to its capacity $C_{\tilde{e}}$, the minimum value of $ \sum_e C_e \phi_e \left(\omega_e^{(N)} \right) $ is the optimal value of the following optimization problem: 
\begin{align*}
    \min_{x\ge 0} \quad &\sum_{e\in \mathcal{E}_{\tilde{e}}}\phi_{e}\left(\omega_e^{(N)} \right) + \phi_{\tilde{e}}(C_{\tilde{e}}) \\
    s.t. &\sum_{j:e\in P_j} x_j = \omega_e^{(N)}, \forall e\in \mathcal{E}_{\tilde{e}}, \\
    \quad &\sum_{j\in I_{\tilde{e}}} x_j = C_{\tilde{e}},
\end{align*} 
where $ x_j, j\in[m] $ is the amount of rate allocated to the $j$th request that uses edge $ \tilde{e} $, $ \mathcal{E}_{\tilde{e}} $ is the set of edges that can share a path of length $m$ with edge $ \tilde{e} $, and $ I_{\tilde{e}} $ is the set of requests that utilizes edge $ \tilde{e} $.
\end{corollary}

\begin{corollary}\label{corol: cr-overall}
The competitive ratio is the maximum of the ratio for the case with edge saturation and the ratio for the case without edge saturation.
\end{corollary}

Assume that $ M $ is a multiple of $m$.
A commonly-employed lower bound of the algorithm output is the final total prices of all edges in the network as follows.
\begin{lemma}\label{lemma:alg-lower-bound}
    $ \textsf{ALG}(\tilde{I}_{j}) \ge \frac{\sum_{e\in \tilde{\mathcal{N}}_j} C_e \lambda_e^{(N)}}{2\gamma(e-1)} $.
\end{lemma}
\begin{proof}
We have
    \begin{align*}
    \textsf{ALG}(\tilde{I}_{j}) &= \sum_{i}v_i/2 + \sum_{i}v_i/2 \\
    &\ge \sum_{i} \frac{r_i}{2} \sum_{e\in P_i} \lambda_e^{(i-1)} + \sum_{i} \frac{r_i d_i}{2} \\
    &= \sum_{i} \frac{r_i}{2}\sum_{e\in P_i} \left(\lambda_e^{(i-1)} + 1 \right) \\
    &= \sum_{i} \sum_{e\in P_i} \frac{r_i \left(\lambda_e^{(i)} - \lambda_e^{(i-1)} \right)}{2[\exp(\frac{r_i \gamma}{C_e})-1]} \\
    &\ge \sum_i \sum_{e\in P_i} \frac{C_e(\lambda_e^{(i)}-\lambda_e^{(i-1)})}{2\gamma (e-1)} \\
    &= \frac{\sum_e C_e \lambda_e^{(N)}}{2\gamma(e-1)},
\end{align*}
where the last inequality follows from that $\frac{x}{e^{\gamma x}-1}$ is decreasing in $ x\in [0,\frac{1}{\gamma}] $.
\end{proof}

It then remains to bound the optimal revenue and final edge prices $ \lambda_e^{(N)} $ for the case with edge saturation, which are dependent on the network topology and the edge capacities.

If $ \forall e, C_e = C $, for a $3M$-long line network, the worst-case scenario is that each almost-saturated edge blocks an $M$-long request with the highest valuation. There are at most 3 such long and valuable requests blocked by three almost-saturated edges. Thus, the optimal revenue is upper-bounded by $3M\bar{p}C$, the online revenue is lower-bounded by $$ \frac{3(C\phi(C-\epsilon) + C(2m-1)\phi((C-\epsilon)/2))}{2\gamma (e-1)}, $$
and the ratio $ \frac{\textsf{OPT}(I_{j})}{\textsf{ALG}(\tilde{I}_{j})} $ is upper-bounded by $ \frac{2(e-1)\gamma M\bar{p}}{\phi(C)+(2m-2)\phi(C/2)} $.

The overall competitive ratio is then upper-bounded by 
$
\max\{4(e-1)\gamma + 2, \frac{2(e-1)\gamma M \bar{p}}{\phi(C) + (2m-2)\phi(C/2)}\}
= \max\{4(e-1)\gamma + 2, \frac{2(e-1)\gamma M \bar{p}}{e^{\gamma} + (2m-2)e^{\gamma/2} + 2m - 1} \} = O(\ln \frac{M\bar{p}}{m})
$
by choosing $ \gamma = O(\ln \frac{M\bar{p}}{m}) $.

For line networks with heterogeneous capacities, denote the minimum capacity in $ \mathcal{N}_j $ as $ \underline{C}_j $ and the edge as $ \underline{e}_j $ and the maximum capacity of edges within $m$ reach of edge $ \underline{e}_j $ in $ \mathcal{N}_j $ as $ \bar{C}_j $. The worst-case scenario happens when edge $ \underline{e}_j $ is almost-saturated and blocks an $M$-long request. The optimal revenue is upper-bounded by $ \sum_{k=j-1}^{j+1} M\bar{p}\underline{C}_k $, the online revenue is lower-bounded by 
$$ \frac{\sum_{k=j-1}^{j+1} \underline{C}_k(e^\gamma -1) + 2(m-1)\bar{C}_k (e^{\frac{\underline{C}_k}{2\bar{C}_k}}-1)}{2\gamma (e-1)}, $$
and the ratio $ \frac{\textsf{OPT}(I_{j})}{\textsf{ALG}(\tilde{I}_{j})} $ is upper-bounded by $ \frac{2(e-1)\gamma M\bar{p}}{e^\gamma - 1 + 2(m-1)\beta (e^{\frac{\gamma}{2\beta}}-1)} $, where $ \beta = \max_j \frac{\bar{C}_j}{\underline{C}_j} $. Thus, the overall competitive ratio is upper-bounded by $ \max\{ 4(e-1)\gamma + 2, \frac{2(e-1)\gamma M\bar{p}}{e^\gamma - 1 + 2(m-1)\beta (e^{\frac{\gamma}{2\beta}}-1)} \} = O(\beta \ln (\frac{M\bar{p}}{2m \beta}+1)) $ by choosing $ \gamma = O(\beta \ln (\frac{M\bar{p}}{2m \beta}+1)) $. When $ \beta $ is large, $ \frac{2(e-1)\gamma M\bar{p}}{e^\gamma - 1 + 2(m-1)\beta (e^{\frac{\gamma}{2\beta}}-1)} $ is upper-bounded by $ \frac{2(e-1)\gamma M\bar{p}}{e^\gamma - 1} = O(\ln(M\bar{p})) $ by choosing $ \gamma = O(\ln(M\bar{p})) $. Thus, the competitive ratio is upper-bounded by $ \max\{O(\ln M\bar{p}), O(\beta \ln (\frac{M\bar{p}}{2m \beta}+1))\} $. 

\end{proof}

\begin{remark}

When edge capacities are identical ($ \beta = 1 $), \textsf{PPM-PS}$_{\phi}$ is $O(\ln{\frac{M\bar{p}}{m}}) $-competitive, recovering the result in~\cite{sun2022online}. When $ \beta = \infty $ and $ m = 1 $, \textsf{PPM-PS}$_{\phi}$ is $O(\ln M\bar{p}) $-competitive, consistent with the result in~\cite{yang2021competitive}.

\end{remark}



\subsubsection{Tree Network}

We consider a directed full binary tree of depth $ M $ and the following two typical arrival patterns in tree networks.

\begin{itemize}
    \item \textsf{Start from Root (SR)}: All requests start from the root node. This corresponds to the case when there is one source node in the communication network.
    \item \textsf{End at Leaf (EL)}: Requests must end at any leaf node but not necessarily start from the root node. This corresponds to the case when every non-leaf node can be the source node and generate data.
\end{itemize}
    
\begin{theorem}\label{thm:tree-net}
For \textsf{SR} requests, when edge capacities are identical (uniform capacity case), the competitive ratio is $ O(2^{m-1}\ln(\frac{M\bar{p}}{m\cdot 2^{m-1}} + 1)) $.
When the $i$-level edges have capacity $\frac{C}{2^i}$ (exponentially-decreasing capacity case), the competitive ratio is $ O(\max\{ \max_{x \in [0,m-1]} 2^x \ln(\frac{M\bar{p}}{m 2^x}+1), \ln(M\bar{p}+1) \}) $.
\end{theorem}

\begin{proof}
    We start with the case when all edges have the same capacity $ C $.
    From Lemma~\ref{lemma:alg-lower-bound}, we have $\textsf{ALG} \ge \frac{\sum_e C \phi_e(\omega_e^{(N)})}{2(e-1)\gamma}$. The optimal revenue is upper-bounded by allocating the capacity of the top-level edge to the highest-valued requests: $ \textsf{OPT}\le M \bar{p} C $. If there is an edge that reaches the capacity, it will be the top-level edge, and there are at most $2^k$ edges at level $k$ among which the load is evenly distributed, then we have
    \begin{align*}
        \sum_e C \phi_e(\omega_e^{(N)}) &\ge C \left[\phi(C) + \sum_{k=1}^{m-1} 2^k \phi\left(\frac{C}{2^k} \right) \right] \\
        &= C \sum_{k=0}^{m-1} 2^k \phi \left(\frac{C}{2^k} \right) \\
        &= C\sum_{k=0}^{m-1} 2^k (e^{\frac{\gamma}{2^k}}-1).
    \end{align*}
    The ratio for the case with edge saturation is then upper-bounded by $ \frac{2(e-1)\gamma \bar{p} C M}{C\sum_{k=0}^{m-1} 2^k (e^{\frac{\gamma}{2^k}}-1)} \le \frac{2(e-1)\gamma \bar{p} M}{m 2^{m-1} (e^{\frac{\gamma}{2^{m-1}}}-1)} $, and the competitive ratio is eventually bounded by $ O(2^{m-1}\ln(\frac{M\bar{p}}{m\cdot 2^{m-1}} + 1)) $ by choosing $ \gamma = O(2^{m-1}\ln(\frac{M\bar{p}}{m\cdot 2^{m-1}} + 1)) $.

    We then consider the influence of heterogeneous capacities on the performance of our mechanism by studying the case where edges at the $i$th level are endowed with a capacity of $\frac{C}{2^i}$. In this case, edges of the first level may not be the first saturated. Let the first saturated edge be of the $l$th level ($l \le m$) and $ \phi_k(\omega) = e^{\frac{2^k \gamma \omega}{C}}- 1 $, the algorithmic output is lower-bounded in
    \begin{align*}
        2(e-1)\gamma \textsf{ALG} &\ge \sum_{k=0}^l \frac{C}{2^k}\phi_k(\frac{C}{2^l}) + \sum_{k=l+1}^m \frac{C}{2^k} 2^{k-l} \phi_k(\frac{C}{2^k}) \\
        &= \sum_{k=0}^l \frac{C}{2^k} (e^{\gamma 2^{k-l}}-1) + \frac{C}{2^l}\sum_{k=l+1}^{m} (e^{\gamma}-1) \\
        &\ge \frac{l\cdot 2^l (e^{\frac{\gamma}{2^l}}-1)+(m-l+1)(e^\gamma - 1)}{2^l} \\
        &\ge \max\{ l\cdot (e^{\frac{\gamma}{2^l}}-1), \frac{(m-l+1)(e^\gamma - 1)}{2^l} \}.
    \end{align*}
    The above inequalities hold for the following reasons. Saturating an edge of the $l$th level (with capacity $\frac{C}{2^l}$) leads to that each of the $l-1$ ancestor edges is consumed at least $\frac{C}{2^l}$, and edges of $ m-l $ child levels are also saturated because each child edge only has one ancestor edge and the total capacities over an edge's child edges are the same to its own capacity. The optimal value is upper-bounded as $\textsf{OPT} \le \frac{M\bar{p}C}{2^l} $ because the bottleneck link is of capacity $\frac{C}{2^l}$. 

    The competitive ratio is then upper-bounded by $ O(\max\{ 2^l \ln(\frac{M\bar{p}}{m 2^l}+1), \ln(\frac{M\bar{p}}{m-l+1}+1) \}) $. We thus complete the proof of Theorem \ref{thm:tree-net}.
    \end{proof}

\begin{theorem}
\label{thm:tree-net-el}
    For \textsf{EL} requests, when edge capacities are identical, the competitive ratio is $ O( \max_{x\in [m,M]} 2^{x-1} \ln (\frac{M\bar{p}}{x 2^{x-1} } + 1 ) ) $. When the $i$-level edges have capacity $\frac{C}{2^i}$, the competitive ratio is $ O(\ln(1+ \frac{M \bar{p}}{m} )) $.
\end{theorem}

\begin{proof}
In a full binary tree, the source nodes of concern must be of depth $d \le M - m$. For identical capacity case, the optimal value is upper-bounded by $ M\bar{p} C $.  Denote the saturated edge with the lowest level as $ \underline{e} $ and the level of its parent node as $\underline{d}$. It is obvious that any ancestor edge of $ \underline{e} $ is also saturated.
The algorithm output $\textsf{ALG}$ is lower-bounded by
\begin{align*}
&\frac{ C\phi(C)+C\sum_{i=1}^{M-\underline{d} - 1}2^i\phi(\frac{C}{2^i}) }{ 2(1-e)\gamma } \\
&= \frac{ C \sum_{i=0}^{M-\underline{d} - 1} 2^i(e^{\gamma/2^i}-1) }{ 2(1-e)\gamma } \\
&\ge \frac{C (M-\underline{d} ) 2^{M-\underline{d} - 1} (e^{\gamma/2^{M-\underline{d} - 1}}-1)}{2(1-e)\gamma}.
\end{align*}

The competitive ratio is upper-bounded by
\begin{align*}
&\max\{ 4(e-1)\gamma+2, \frac{2(e-1)M\bar{p}\gamma}{(M-\underline{d} ) 2^{M-\underline{d} - 1} (e^{\gamma/2^{M-\underline{d} - 1 } } - 1)} \} \\
&\le O( 2^{M-\underline{d} -1} \ln (\frac{M\bar{p}}{(M-\underline{d}) 2^{M-\underline{d}-1} } + 1 ) ),
\end{align*}
by choosing $\gamma = O( 2^{M-\underline{d} -1} \ln (\frac{M\bar{p}}{(M-\underline{d}) 2^{M-\underline{d}-1} } + 1 ) ) $. Thus, the competitive ratio is upper-bounded by $ O( \max_{x\in [m,M]} 2^{x-1} \ln (\frac{M\bar{p}}{x 2^{x-1} } + 1 ) ) $.

For the case where the capacity of the $ i $-th level edge is $ \frac{C}{2^i} $, the optimal value is upper-bounded by $ M \bar{p} \frac{C}{2^{\underline{d}}} $, where $ \underline{d} $ is defined the same as above. The algorithmic output is lower-bounded in $ 2(e-1)\gamma \textsf{ALG} \ge \frac{C}{ 2^{\underline{d}} } \sum_{k=0}^{M-\underline{d}-1} \frac{C}{2^k} 2^k (e^\gamma - 1) $. The ratio for the saturated case is thus upper-bounded by $ \frac{2(e-1)\gamma  M \bar{p} \frac{C}{2^{\underline{d}}}}{\frac{C}{ 2^{\underline{d}} } \sum_{k=0}^{M-\underline{d}-1} \frac{C}{2^k} 2^k (e^\gamma - 1)} = \frac{ 2(e-1)\gamma M \bar{p} }{ (M-\underline{d})(e^\gamma - 1) } = O(\ln(1+ \frac{M \bar{p}}{M-\underline{d}} )) $ by choosing $ \gamma = O(\ln(1+ \frac{M \bar{p}}{ M-\underline{d} } )) $. Thus, the competitive ratio is upper-bounded by $ O(\ln(1+ \frac{M \bar{p}}{m} )) $.

\end{proof}

\begin{remark}
The following observations and understandings are gained:
    \begin{itemize}
        \item For \textsf{SR} requests, the competitive ratio for uniform capacity case is smaller than that for the exponentially-decreasing capacity case, demonstrating the difficulty of the latter.
        \item The \textsf{EL} requests are harder than \textsf{SR} requests when edge capacities are identical, while it is the opposite for exponentially-decreasing capacity case. It shows the intricate interaction between the request pattern and the network design, especially the configuration of link capacities.
        \item We also observe that the competitive ratio of \textsf{EL} requests with exponentially-decreasing capacity is the same as that of the line network with uniform capacity. The matching between the request pattern and the network configuration should thus be deemed as vital for a good competitive ratio.    
    \end{itemize}
\end{remark}

\begin{remark}
    Theorem~\ref{thm:tree-net} and Theorem~\ref{thm:tree-net-el} can be extended trivially to the case when the branching factor of the tree is $b>2$.
\end{remark}

\subsection{Main Results II: Impact of System-Level Costs on Optimal Design of \textsf{PPM-PS}$_{\phi}$}

As mentioned in the Introduction, the incurred waiting time at nodes leads to a system-level cost, namely the service quality degradation caused by the network congestion, requiring additional consideration in balancing between revenues and costs to further improve social welfare. We model the cost as follows. The sum of mean rates for all flows passing through edge $e$ is defined as $\omega_e = \sum_{i\in [N]} r_i \delta_{i}^e $, where $\delta_{i}^e=1$ indicates that edge $e$ is on the path of $i$th agent $P_i$, and $\delta_{i}^e = 0$ indicates the opposite. 
In this work, each edge is modeled as an $M/M/1$ queue with the arrival rate $\lambda_e$ and the service rate $C_e$. The cost is quantified by the total number of packets in the network $\sum_{e} f(\rho_e)$, where 
$$
f(\rho_e)=
\begin{cases}
\frac{\rho_e}{1-\rho_e}, & 0 \le \rho_e < 1,\\
\infty,                  & \rho_e \ge 1,
\end{cases}
$$
and $ \rho_e = \frac{\omega_e}{C_e}$ is the utilization of edge $e$. It is well-accepted in the queuing theory that the number of packets in the network is positively related to the average network delay, and thus a useful indicator for the network congestion. Other typical preferences of network operators, such as maximizing the minimum load, can also be incorporated by enforcing different $f$'s. In this regard, $f$ can be viewed as a regularizer of the network state. 

When the minimum path length $m>1$, the worst-case instance is topology-dependent as shown in the previous subsection. To avoid over-complicating the problem and enable the analysis for the case with costs, we consider $m=1$ here. 

Theorem~\ref{theorem_sufficiency} provides a set of sufficient conditions on the pricing function $\phi$ to remain competitive after considering the system-level cost. 

\begin{theorem}[Sufficiency] \label{theorem_sufficiency}
For any given $\gamma\ge 1$, $\textsf{PPM-PS}_{\bm{\phi}}$ is $\alpha$-competitive if $ \bm{\phi} = (\phi_e)_{\forall e\in \mathcal{E}} $ and $ \phi_e: [0,\bar{\rho}_e] \rightarrow \mathbb{R}^+ $ is an analytic and non-decreasing solution to the following differential equation with boundary conditions:
\begin{align}\label{eq_BVP_sufficiency}
\begin{cases}
\left( 1 - {(C_e\phi_e)^{-1/2}} \right) \phi'_e = \gamma\left(\phi_e-f'/C_e\right) ,\\
\phi_e(0) = \frac{1}{C_e}, \phi_e(\bar{\rho}_e) \ge \bar{p},
\end{cases}
\end{align}
where $ \bar{\rho}_e $ satisfies $ f'(\bar{\rho}_e) = \bar{p} C_e $.
\end{theorem}

\begin{proof}
    Before any agent arrives, $ \textsf{ALG} = 0$, and the optimal $\textsf{OPT}$ is upper-bounded by $ \sum_{e}f^*(\phi_e(0)C_e) \ge 0$, where $f^*(y)= \sup_{\rho \in [0,\infty)} [ y\rho - f(\rho) ] 
    = \begin{cases}
        (\sqrt{y}-1)^2,   \text{if } y \ge 1, \\
        0, \text{if }  y < 1.
    \end{cases}$
    By setting $\phi_e(0) = \frac{1}{C_e}$, we have $\textsf{ALG} = \textsf{OPT} = 0$.

     If agent $i$ joins the system, the social welfare increases by

    \begin{align}
    \nonumber \Delta \textsf{ALG}
    & = v_i - \sum_{e\in P_i} \left[f (\rho_e^{(i)}) - f (\rho_e^{(i-1)})\right] \\
    \nonumber & = \mu_i + \sum_{e\in P_i} r_i p_e^{(i-1)} - \sum_{e\in P_i} \left[ f(\rho_e^{(i)}) - f(\rho_e^{(i-1)}) \right],
\end{align}
where $ \mu_i $ is the value margin of agent $i$, i.e., $ \mu_i = v_i - r_i \lambda^{(i-1)(P_i)} \ge 0 $ and $ \rho_e^{(i)} = \frac{\omega_e^{(i)}}{C_e} $.

The optimal social welfare increases at most
\begin{align}
    \nonumber D_i-D_{i-1} = \mu_i + \sum_{e\in P_i} \left[ f^*(\lambda_e^{(i)} C_e) - f^*(\lambda_e^{(i-1)} C_e) \right].
\end{align}

Being $\gamma$-competitive is implied by the following inequality:
\begin{align}
    \nonumber \sum_{e\in P_i} r_i \lambda_e^{(i-1)} - \sum_{e\in P_i} \left[ f (\rho_e^{(i)}) - f(\rho_e^{(i-1)}) \right] + \left(1-\frac{1}{\gamma} \right)\mu_i \\
    \nonumber \ge \frac{1}{\gamma} \sum_{e\in P_i} \left[ f^*(\lambda_e^{(i)} C_e) - f^*(\lambda_e^{(i-1)} C_e) \right],
\end{align}
which is further implied by the inequality over each agent as follows:
\begin{align}
\label{eq:individual-ineq}
     \nonumber r_i \lambda_e^{(i-1)} - \left[ f (\rho_e^{(i)}) - f(\rho_e^{(i-1)}) \right] \\
     \nonumber \ge \frac{1}{\gamma} \left[ f^*(\lambda_e^{(i)}C_e) - f^*(\lambda_e^{(i-1)}C_e) \right].
\end{align}
By dividing $r_i = C_e ( \rho_e^{(i)}-\rho_e^{(i-1)} )$ at both sides, we have
$
    \forall e\in P_i, \quad \lambda_e^{(i-1)} - \frac{1}{C_e}\frac{f(\rho_e^{(i)}) - f( \rho_e^{(i-1)} )}{\rho_e^{(i)}-\rho_e^{(i-1)}} \ge \frac{1}{\gamma } \cdot \frac{f^*(\lambda_e^{(i)} C_e) - f^*(\lambda_e^{(i-1)} C_e)}{ (\lambda_e^{(i)} - \lambda_e^{(i-1)}) C_e } \cdot \frac{\lambda_e^{(i)} - \lambda_e^{(i-1)}}{\rho_e^{(i)}-\rho_e^{(i-1)}},
$ which is implied by the differential equation
$$
\phi_e\left(\rho_e^{(i-1)}\right) - \frac{1}{C_e} f'_e\left( \rho_e^{(i-1)} \right) \ge \frac{1}{\gamma} f^{*'} ( \phi_e (\rho_e^{(i-1)}) C_e ) \phi'_e(\rho_e^{(i-1)}).
$$
The right boundary condition is determined by identifying the worst-cast instance and ensuring the $\gamma$-competitiveness for it. One worst-case instance consists of agents who can be routed via unit-length paths. Two phases in this instance exist: in the first phase, there come agents with valuation increasing from $1$ to $\bar{p}$; followed by agents with valuation $ \bar{p}-\epsilon $ in the second phase.
\end{proof}

Theorem~\ref{theorem_necessity} reinforces the significance of Eq.~\eqref{eq_BVP_sufficiency} by showing that when the rate-to-capacity ratio $\epsilon$ is infinitesimal, the existence of a $\gamma$-competitive online mechanism for the case with cost is equivalent to the existence of a solution to the integral version of Eq.~\eqref{eq_BVP_sufficiency}.

\begin{theorem}[Necessity] \label{theorem_necessity}
For any $\gamma>0$, if there exists an $ \gamma $-competitive deterministic online mechanism (not necessarily PPMs) for the online path selection problem with convex costs and the infinitesimal rate-to-capacity ratio, then the integral version of Eq.~\eqref{eq_BVP_sufficiency} has at least one solution.
\end{theorem}

\begin{proof}
    Denote a group of agents with value density $\nu$ and total demand $ C_e f^{*'}(\nu C_e) $ as $G_\nu$. Consider the following instance $I_p$ indexed by $p$, $ p\in [0, \bar{p})$: there come $G_\nu$s with $\nu$ increasing from $0$ to $p$ continuously. After that, there comes $G_\nu$ with $\nu=p-\epsilon$. The optimal solution is composed of all agents in the last group of the instance $I_p$, i.e., group $G_{p-\epsilon}$, and its welfare is 
    $
    (p-\epsilon) C_e f^{*'}((p-\epsilon) C_e) - f(f^{*'}((p-\epsilon) C_e)) = f^*((p-\epsilon)C_e).
    $
    Define the utilization of link $e$ of any $\alpha$-competitive online algorithm after processing $G_\nu$ as $\psi_e(\nu)$. Denote the output of an online algorithm as $\textsf{ALG}$. Given the $\alpha$-competitiveness, the following inequality holds for $\forall p\in \left(1/C_e, \bar{p}\right]$:
\begin{align}
    \nonumber \textsf{ALG} &= \int_{1/C_e}^{p} \nu C_e d\psi_e(\nu) - f(\psi_e(p)) \\
    &\ge \frac{1}{\gamma} \textsf{OPT} = \frac{1}{\gamma} f^*((p-\epsilon)C_e).
\end{align}

If there exists a $\gamma$-competitive online algorithm (not necessarily PPM), we can always construct a PPM with $ \psi_e(\bar{p}) = \bar{\rho}_e $ and $\psi_e(\frac{1}{C_e})=0$ to be at least $\gamma$-competitive. 

The construction is as follows: Due to the definition of $\psi_e(p)$, we have $\psi_e(\nu)\ge \psi_e(\frac{1}{C_e}) \ge 0$, for all $\nu\in [\frac{1}{C_e},\bar{p}]$. If $\psi_e(\frac{1}{C_e}) > 0$, agents that incur negative welfare ($v_i/r_i \le \frac{1}{c_e} $) will join, and thus there always exists a more competitive online algorithm with $\psi_e(\frac{1}{c_e})=0$. If $\psi_e(\bar{p}) > \bar{\rho}_e$, we can always construct an algorithm at least $\gamma$-competitive by stopping the allocation right before the utilization hits the effective utilization $\bar{\rho}_e$, because the increase of the link costs after exceeding the effective utilization is greater than the increase of the value; if $\psi_e(\bar{p})<\bar{\rho}_e$, we can always allocate the remaining $\bar{\rho}_e - \psi_e(\bar{p})$ of link $e$ to $I_{\bar{p}}$ and achieve a competitive ratio no worse than $\gamma$.

Thus, for any $\gamma$-competitive online algorithm, there is a $\psi_e$ that satisfies Eq.~\eqref{eq:lower-bound-tight}:
\begin{align}
    \label{eq:lower-bound-tight}
    \begin{cases}
        \int_{1/C_e}^{p} \nu d\psi_e(\nu) - \frac{1}{C_e}f(\psi_e(p)) \ge \frac{1}{\gamma C_e}f^*(p C_e), \forall p\in (1/C_e, \bar{p}) \\
        \psi_e(\frac{1}{C_e})= 0, \psi_e(\bar{p}) = \bar{\rho}_e.
    \end{cases}
\end{align}

Define $\underline{\psi_e}(\nu)$ as the infimum over all feasible solutions to Eq.~\eqref{eq:lower-bound-tight}: $ \underline{\psi_e}(\nu) = \inf \{ \psi_e(\nu) | \psi_e$ is non-decreasing and feasible for Eq.~\eqref{eq:lower-bound-tight}$\}$. One can show that $\underline{\psi_e}$ is feasible for Eq.~\eqref{eq:lower-bound-tight} with the equality holds and is strictly increasing.

Construct $\varphi_e$ as follows: for any $p\in (\frac{1}{C_e},\bar{p})$, $ \varphi_e(\rho) = \underline{\psi_e}^{-1}(\rho) = p, \forall \rho \in (0,\bar{\rho}_e)$, $\varphi_e(0) = \frac{1}{C_e}$, $\varphi_e(\bar{\rho}_e) = \bar{p}$. By replacing $ \nu $ with $ \varphi_e(s) $ in Eq.~\eqref{eq:lower-bound-tight}, we have
$
    \int_{0}^{\rho} \varphi_e(s) ds - \frac{1}{C_e}f(\rho) = \frac{1}{\alpha C_e}f^*(\varphi_e(\rho) C_e), \forall \rho \in (0, \bar{\rho}_e),
$
which shows that $\varphi_e$ is a solution to the following set of equation:
\begin{align}
\label{eq:integral-ineq-infinitesimal}
    \nonumber &\int_0^\rho \phi_e(s) ds - \frac{1}{C_e}(f(\rho) - f(0)) \\
    \nonumber &\ge \frac{1}{\gamma C_e} \cdot \left( f^*(\phi_e(\rho)C_e ) - f^*(\phi_e(0)C_e) \right) \\
    &= \frac{1}{\gamma C_e} \cdot f^*(\phi_e(\rho)C_e ),
\end{align}
which is the integral version of Eq.~\eqref{eq_BVP_sufficiency}.
\end{proof}


\section{Experiments}\label{sec: experiments}

\subsection{Description of Experiment Setting}
\begin{figure}[t!]
    \centering
    \begin{minipage}[b]{\linewidth}
        \centerline{\includegraphics[width=0.7\textwidth]{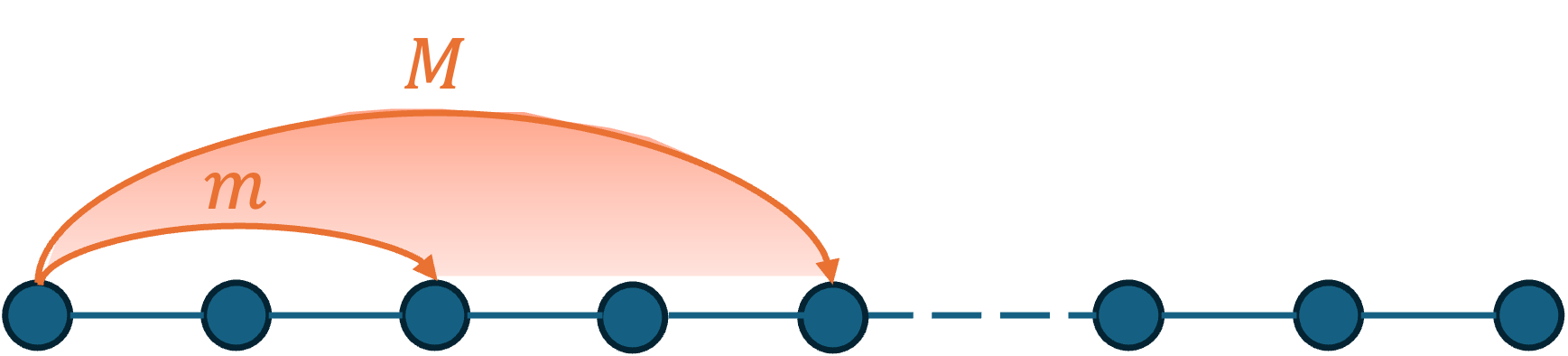} }
        \centerline{(a) Line Network}
        \vspace{1em}
        \centerline{\includegraphics[width=0.7\linewidth]{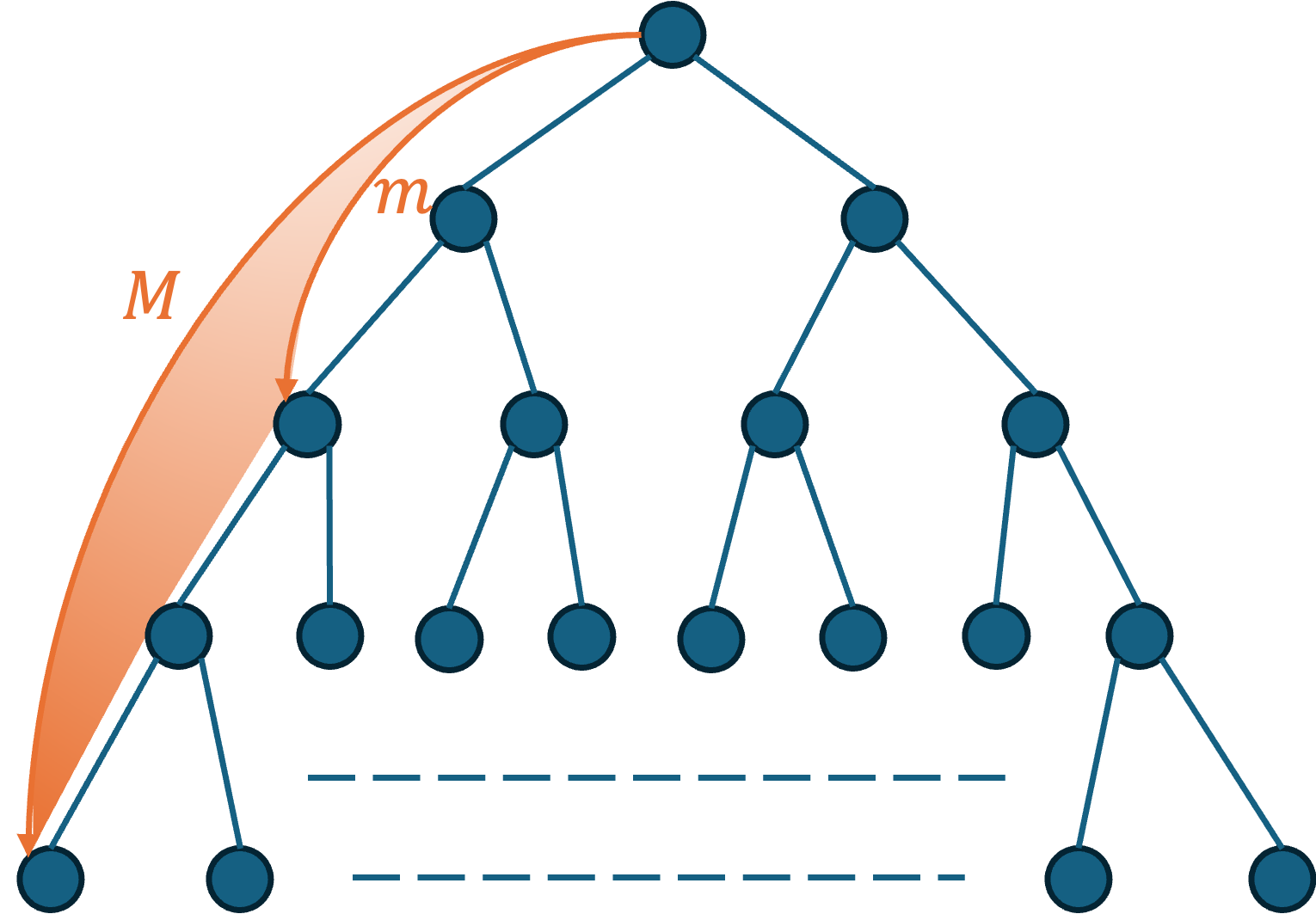}}
        \centerline{(b) Tree Network with SR requests}
    \end{minipage}
    \caption{Illustration of Network Topology}
    \label{fig:topo}
\end{figure}

In line networks, 100 nodes in a line are connected by edges of uniform capacity 100. We conduct $20$ independent simulations, each generating $300$ requests with $ \bar p = 6 $. In particular, the path length and the value density of a request are drawn uniformly random from $[m, M]\cap \mathbb{N}$ and $[1, \bar p]$, respectively. 

In tree networks, a full binary tree of depth $M$ is generated, with the capacity of the two edges closest from the root set at 2560, and capacity of subsequent edges decaying exponentially at each level. We conduct $40$ independent simulations, each generating $3000$ requests that originate from the root with a random trajectory down the tree. Similarly, the path length and the value density of a request are drawn uniformly random from $[m, M]\cap \mathbb{N}$ and $[1, \bar p]$, respectively. Figure~\ref{fig:topo} illustrates the setting.

In both cases, we compute $\textsf{OPT}(I)$ by solving an integer programming problem with Gurobi for each arrival instance $I$.

Then, in an effort to simulate a hard arrival instance that challenges the online algorithm, we construct requests with progressively longer paths and increasing value densities for each given path length. This strategy aims to trick the online algorithm into accepting shorter and lower-valued requests and therefore leading to bottlenecks. In contrast, the optimal strategy in hindsight is to reject such requests and accept requests of higher value and longer length.

Lastly, we run an experiment to examine the relationship between the competitive ratio and the maximum price $\bar{p}$ for the case with cost.

(\textbf{Choices of $\gamma$})
Optimal designs that hedge against worst cases usually exhibit overly cautious and conservative empirical behaviours. To address this issue, many attempts have been made to design data-driven designs that also deliver effective practical performances~\cite{zeynali2021data}. In our design, the larger the $\gamma$, the more rapidly the link price increases and therefore the more conservative the online algorithm is. We conduct experiments to examine how price aggressiveness impacts the empirical performance in different topologies and path length bounds.

\subsection{Experimental Findings and Discussions}
\subsubsection{Impacts of Maximum Path Length}
\begin{figure}[t]
    \centering
    \includegraphics[width=\linewidth]{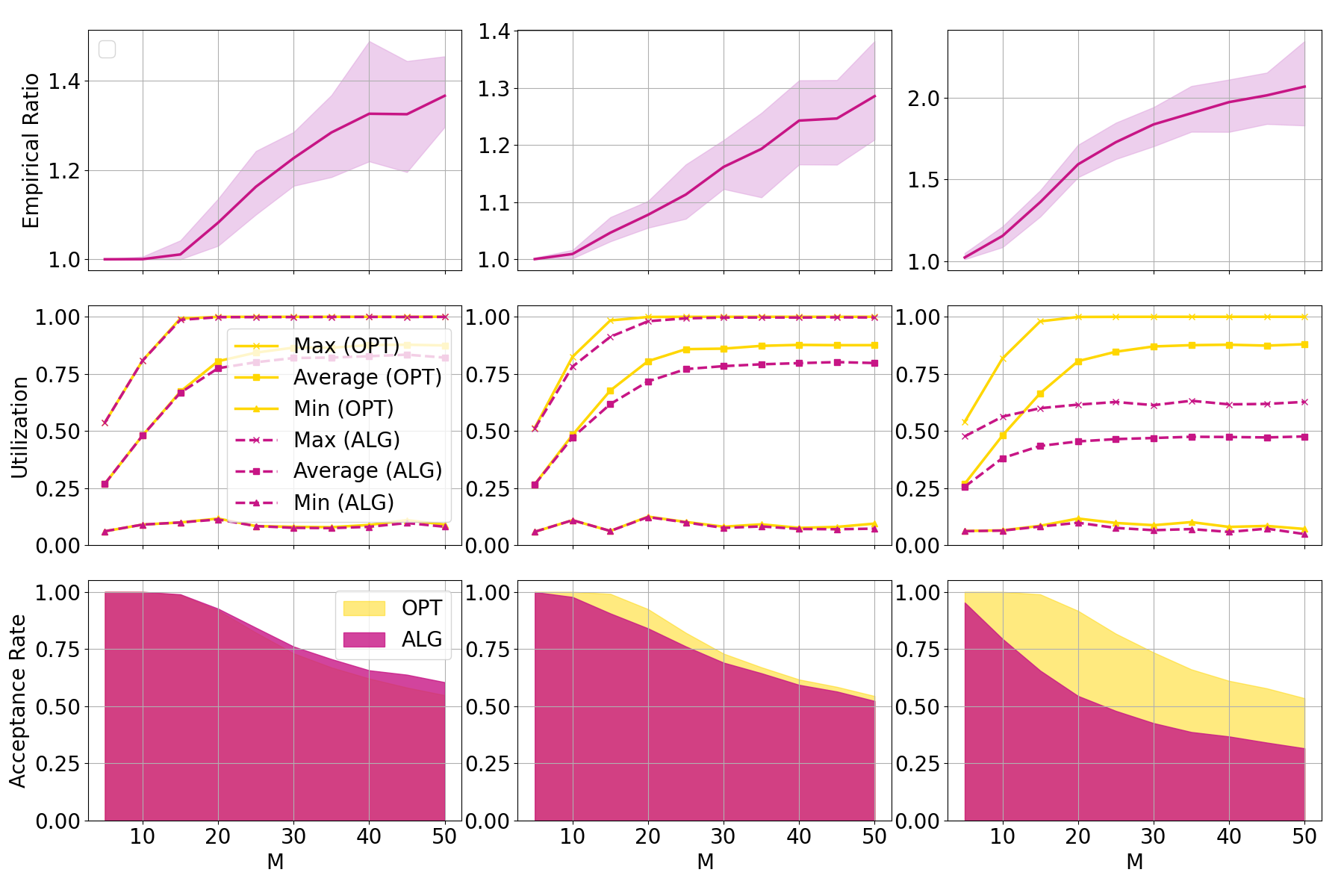}
    \caption{\textbf{Impact of maximum path length $M$ in line networks.} We conduct experiments for different values of $M$ while fixing the minimum path length $m$ at 1 for different values of $\gamma$: 0.5 (left), 2 (middle), and 4 (right). We plot the empirical ratio $\textsf{OPT}(I)/\textsf{ALG}(I)$, utilization statistics of edges in the network, and the acceptance rate of requests.}
    \label{fig:line_max_len}
\end{figure}


(\textbf{Line Network}) In Figure~\ref{fig:line_max_len}, we see that when $\gamma\in \{0.5, 2, 4\}$, the empirical ratio increases in a logarithmic manner with $M$, which is consistent with the competitive ratio. The intuition is that the optimal allocation $\textsf{OPT}$ possesses greater power over choosing requests of higher-valued requests of longer paths as $M$ increases. Additionally, we observe rather consistent increasing trends in edge utilization levels as requests of longer path penetrate further into the line network. Meanwhile, there still exist edges with relatively low utilization levels as the maximum path length is still small compared to the total number of edges in the network. On the other hand, a longer request stands a greater chance of passing through a highly utilized edge, leading to an overall higher price to fulfill this request and thus a drop in the acceptance rate. This decline in the acceptance rate is particularly pronounced when $\gamma=4$ as it leads to the sharpest increase in edge prices with utilization levels due to its implied conservativeness.

\begin{figure}[t]
    \centering
    \includegraphics[width=\linewidth]{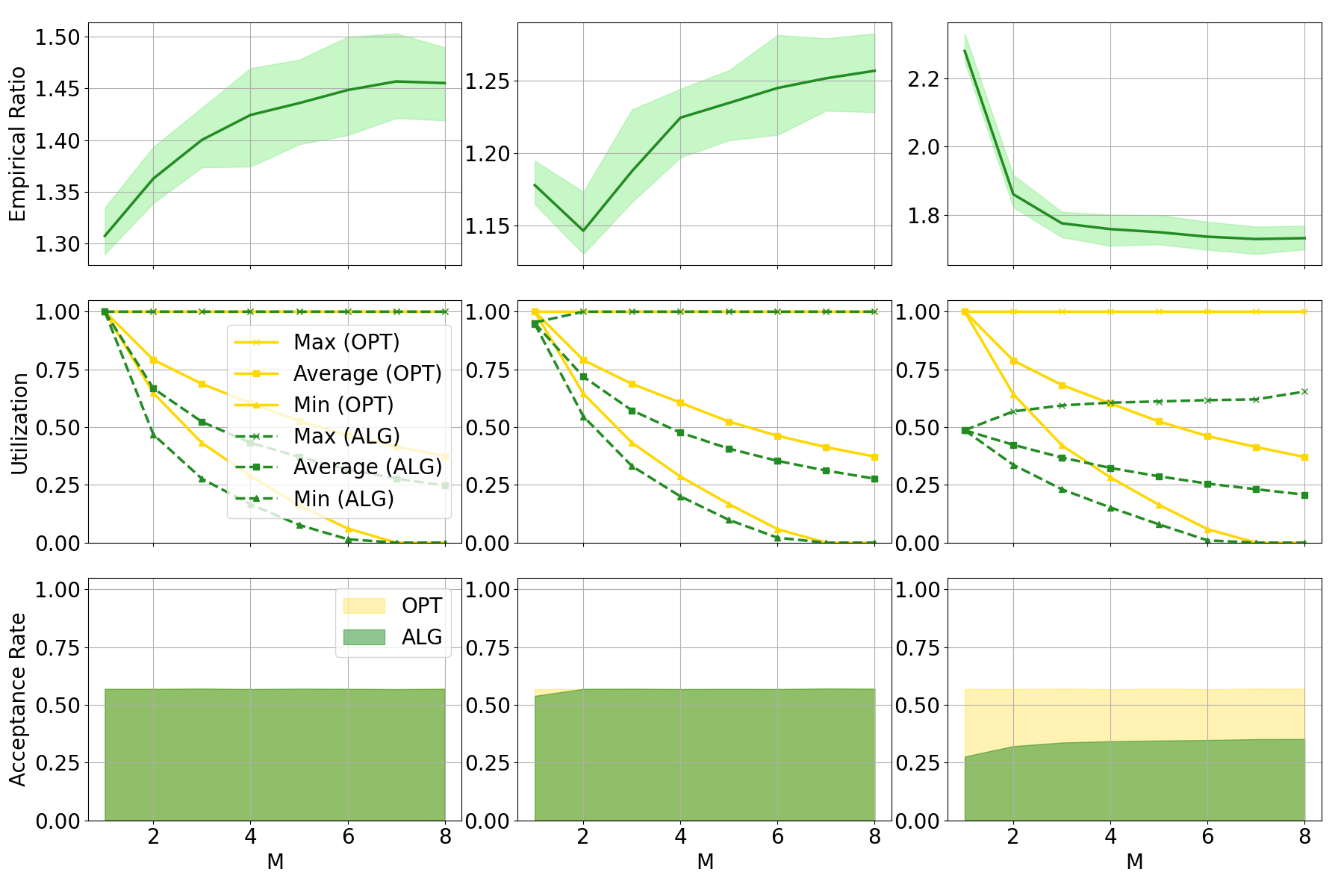}
    \caption{\textbf{Impact of maximum path length $M$ in tree networks.} The minimum path length $m$ is fixed at $1$ for different values of $\gamma$: 0.5 (left), 2 (middle), and 4 (right).}
    \label{fig:tree_max_len}
\end{figure}

(\textbf{Tree Network}) Figure~\ref{fig:tree_max_len} exhibits conflicting relations between $M$ and the empirical ratio.
When $\gamma = 0.5 $, the empirical ratio increases with $M$; when $\gamma = 2$, the empirical ratio also increases with $M$ with a drop when $M = 2$; when $\gamma = 4$, the empirical ratio decreases, which is contrary to our theoretical findings. It sheds light on the gap between the worst-case analysis and the empirical performance.

Firstly, when $\gamma=4$, both the acceptance rate of the online algorithm and the maximum utilization over all edges experience a significant drop compared to the cases when $\gamma\in\{0.5, 2\}$ due to its high price. In this case, the network is far from being saturated and is overall under-utilized. Thus, the worst-case scenario of having bottlenecks due to edge saturation has not come into the picture yet. It is also not hard to understand why the empirical ratio decreases with the maximum length $M$ in the unsaturated case. When $M$ increases, the average value of users increases, which implies that being conservative at larger $M$ values is more advantageous because future agents are more likely to bring a much higher value.

Moreover, the empirical ratio is the smallest when $\gamma=2$, which illustrates that a good $\gamma$ should balance between being overly aggressive ($\gamma=0.5$) and overly conservative ($\gamma=4$), and thus achieve a good trade-off between maximizing the resource utilization efficiency and reserving enough resources for the future. It can be done by increasing $\gamma$ for larger $M$ values while maintaining sufficient utilization levels. In addition to optimizing $\gamma$, it also indicates that the maximum path length has a negative effect in the empirical ratio. We advise that network operators keep this in mind and scale the network without increasing the average path length too much. A possible guideline is to try to avoid deep trees in the network and keep an as-flat-as-possible structure.

\subsubsection{Impacts of Minimum Path Length}

The trends of the empirical ratio with $m$ in line and tree networks and results for hard instances are discussed in the following. In short, we observe interesting trends for stochastic instances that are partially inconsistent with the theoretical results from the worst-case analysis framework and offer detailed analysis for each of them.

\begin{figure}[t]
    \centering
    \includegraphics[width = \linewidth]{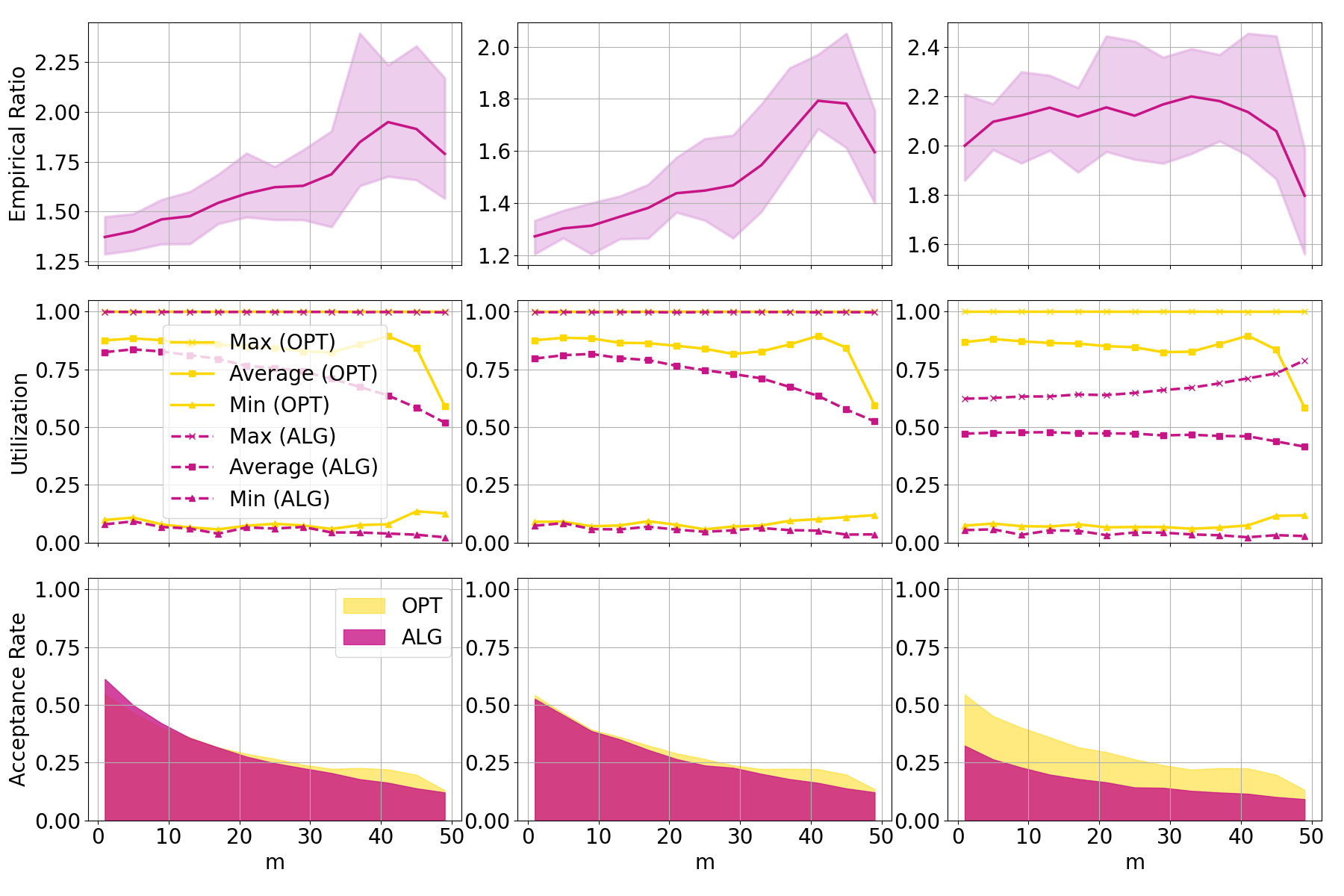}
    \caption{\textbf{Impact of minimum path length $m$ in line networks.} The maximum path length $M$ is fixed at $ 50 $ for different values of $\gamma$: 0.5 (left), 2 (middle), and 4 (right).}
    \label{fig:line_min_len}
\end{figure}

(\textbf{Line Network}) Figure~\ref{fig:line_min_len} shows the relationship between the empirical ratio and the minimum path length $ m $ for line networks under different price aggressiveness $ \gamma \in \{0.5, 2, 4\} $.
As the minimum path length $ m $ increases towards the maximum path length $ M $, the range of possible starting points that can guarantee a minimum path length $ m $ narrows, making it more likely that requests share a common set of edges. These common edges are thus frequently requested and more likely to be rather saturated, hindering the online algorithm from selecting highest-valued requests whereas the optimal allocation can perform the maximization \textit{ex post}.

When there exist saturated edges as shown in the utilization row of Figure~\ref{fig:line_min_len}, as $ m $ increases, in a line network with homogeneous capacities, the neighbouring edges tend to be extensively utilized as well. Therefore, more requests are rejected due to the resulting high price. It is thus better to be somewhat conservative and avoid using up too much capacity too fast by accepting many lower-valued requests. Due to such susceptibility to the saturation, it is thus more challenging for the online algorithm to perform as close as possible to the optimal allocation, leading to an increase in the empirical ratio. 

Interestingly, we also observe a sudden drop in the empirical ratio when $ m $ approaches $ M = 50 $. In this case, requests become increasingly similar in that there are many common edges requested. We conjecture that this results in a significantly reduced ability for even the optimal allocation to optimize, as this phenomenon can be largely explained by the drop in its average utilization rate.



\begin{figure}[t!]
    \centering
    \includegraphics[width = \linewidth]{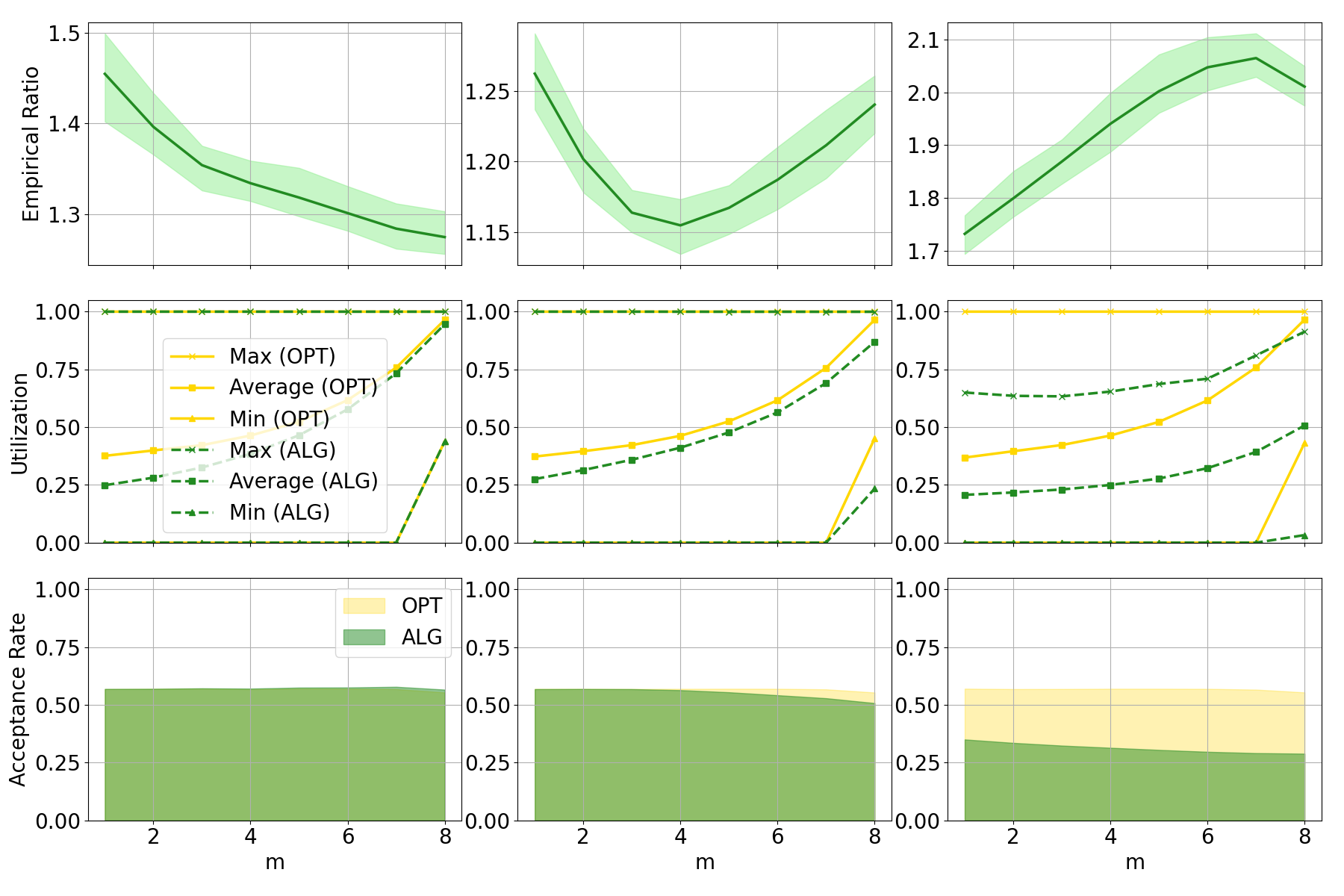}
    \caption{\textbf{Impacts of minimum path length $m$ in tree networks.} The maximum path length $M$ is fixed at $ 8 $ for different values of $\gamma$: 0.5 (left), 2 (middle), and 4 (right).}
    \label{fig:tree_min_len}
\end{figure}

(\textbf{Tree Network}) Figure~\ref{fig:tree_min_len} shows the relationship between the empirical ratio and the minimum path length $ m $ for tree networks under different price aggressiveness $ \gamma \in \{0.5, 2, 4\}$.
We see that when $ \gamma=0.5 $, the empirical ratio decreases with $m$, which is consistent with our theoretical findings. We also observe intricate trends in the empirical ratio not explained by our theoretical results. We argue that this is due to the inherent limitation of the worst-case analysis framework as it cannot effectively characterize empirical performance on stochastic inputs. In particular, when $\gamma = 2$, the empirical ratio decreases at first and then increases with $m$; when $\gamma = 4$, the empirical ratio increases with $m$ with a minor drop when $m=M=8$. We offer discussions on the observed phenomena in the following.

When $m$ increases, the requests will be required to travel through more edges in the network, and thus the average edge utilization levels of both the online algorithm and the optimal allocation increase.
When $m=M=8$, all requests are directed all the way down to leaf nodes. Therefore, the minimum utilization of the optimal allocation becomes positive, as all edges, even at the lowest level, will be requested in expectation.

\begin{figure}[t!]
    \centering
    \includegraphics[width = 0.89\linewidth]{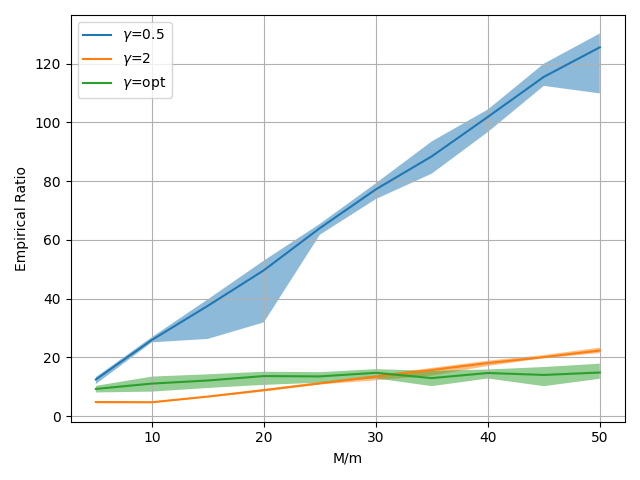}
    \caption{\textbf{Hard instances in line networks}: empirical ratio vs. path length variation $ M/m $. The minimum path length $ m $ is fixed at $ 1 $.}
    \label{fig:line-wc-max-len}
\end{figure}

\begin{figure*}[t!]
    \centering
        \subfigure[Empirical ratio vs. $m$ ($M=8$)]{
            \includegraphics[width = 0.45\linewidth]{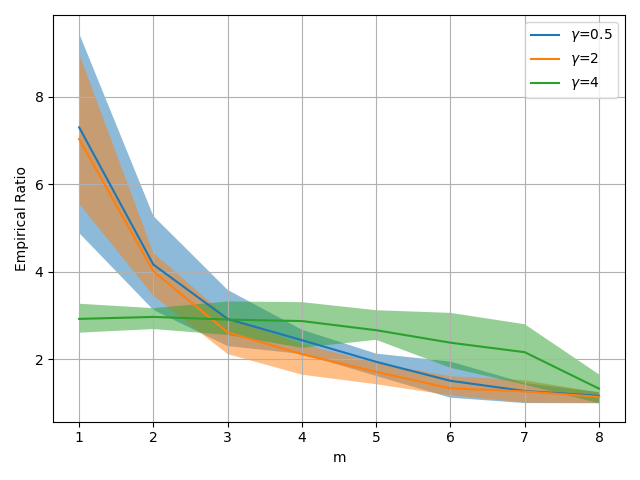}
        }
        \subfigure[Empirical ratio vs. $M$ ($m=1$)]{
            \includegraphics[width = 0.45\linewidth]{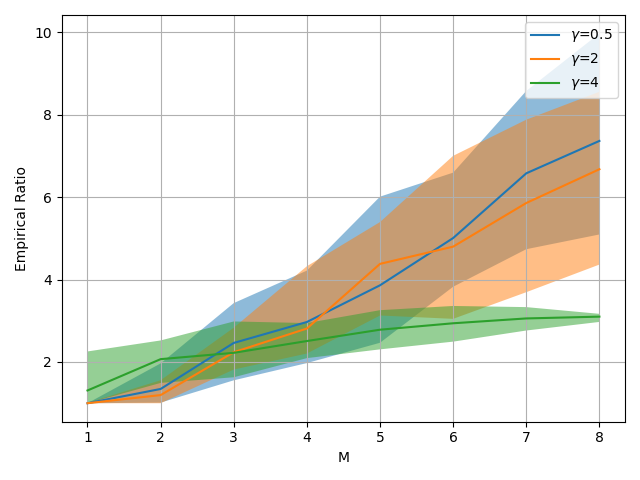}
        }
        \caption{Hard instances in tree networks}
        \label{fig: tree-wc}
\end{figure*}
    
Additionally, we observe that the utilization levels and the acceptance rate of the online algorithm ($\gamma=0.5$) and the optimal allocation are almost identical.
It means that the optimal allocation strategy acts closely to a greedy scheme when the environment is stochastic.
However, in terms of the empirical ratio, the online algorithm ($\gamma=2$) still outperforms the other two.
It clearly shows that an online algorithm does not need to closely mimic the optimal allocation to have a small empirical ratio.
    
It is no surprise that the empirical ratio of the online algorithm ($\gamma=4$) increases with $m$ given that the maximum utilization of the optimal is strictly smaller than $1$.
The instances generated do not incur any edge saturation for any $m$.
It means that in the online algorithm, every request leaves the network because of an insufficient value.
However, with $m$ increasing, it is more likely for two requests to share more edges, leading to more coupling effects between requests.
Consider two requests with path length difference $1$, and one path contains the other.
Assume that the value densities of the two paths are such that the longer one can provide a higher value but the value density of the shorter one is larger.
The longer one can be easily told off if the shorter one is accepted first, and thus it creates a harder time for the online algorithm to distinguish between them.
But it is much easier for the algorithm to drop the ones with a smaller value if $m$ is smaller because it is less likely to encounter the aforementioned conundrum, where one needs to choose between requests with conflicting values and value densities.

When there exists edge saturation (maximum utilization is 1 for $ \gamma \in \{0.5, 2\} $), it verifies the theory that the ratio decreases with $ m $ when there exists edge saturation.
But the theory does not predict as to why the ratio increases (right half of $ \gamma=2 $).
Observing the utilization, as soon as the online algorithm starts to accept less than the optimal, the empirical ratio increases.
One possible reason is that by setting $ \gamma=2 $, there exists a mix of saturated edges and unsaturated edges in the network, and the empirical ratio increases due to the same reason as $ \gamma=4 $.
This reveals again the limitation of the worst-case analysis approach when applied to complex multi-dimensional systems, for which worst-case instances are typically hard to pin down. 

\subsubsection{Hard Instances}
Having observed mixed results against stochastic inputs, we are interested in finding out how the online algorithm performs against challenging instances. In face of strategically constructed hard instances, we observe replicating trends that match our theoretical findings.

The construction of hard instances is as follows: we generate requests with progressively long path length, and for each path length, we gradually feed the online algorithm requests with increasing value densities, starting from the lower bound. In the second part, we provide high-valued requests with long path length. In essence, we want to trick the online algorithm into accepting shorter, lower-valued requests in the first part and create bottlenecks that prevent it from accepting requests in the second part. On the other hand, the optimal allocation strategy should simply ignore requests in the first part and accept requests in the second part.

\begin{figure}[t!]
    \centering 
    {\includegraphics[width=4.9cm]{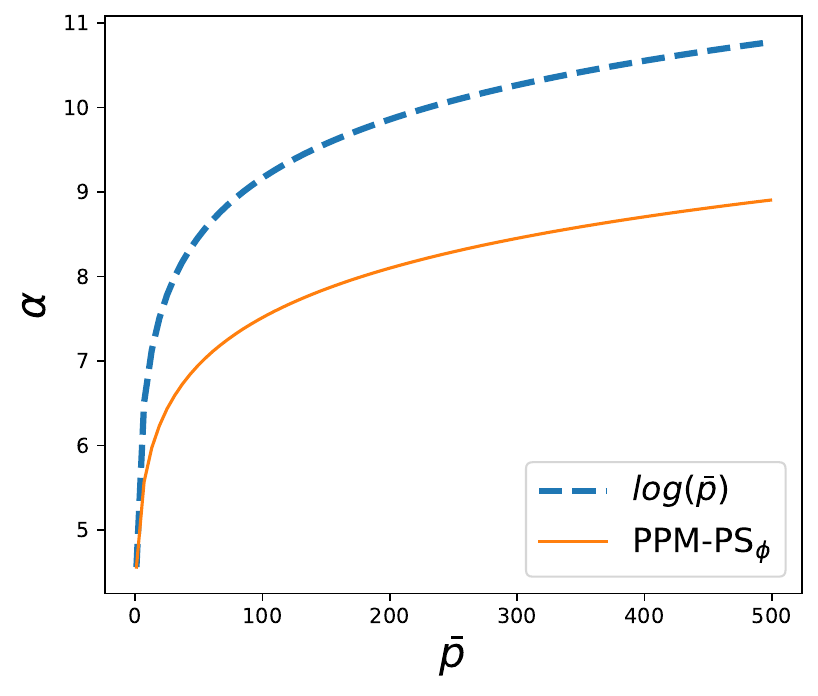}}
    \caption{Best competitive ratio of $\textsf{PPM-PS}_{\bm{\phi}}$ vs. $\bar{p}$.}
    \label{fig:alpha_rho}
\end{figure}

Figure~\ref{fig:line-wc-max-len} shows the relationship between the empirical ratio and $ M/m $ in line networks, where $ M/m $ indicates the fluctuation ratio of path lengths. Intuitively, a larger ratio indicates greater uncertainty and brings additional challenges. As shown before, the order-optimal choice of $ \gamma $ is given in Theorem~\ref{thm: upbds-line}, and we select $ opt = 2\ln(\frac{(e-1)M\bar{p}}{m}+1) $ in the experiment. For $ \gamma \in \{0.5, 2\} $, the empirical ratio increases linearly, whereas when $ \gamma = opt $, the empirical ratio grows logarithmically, reasserting the order optimality of our results.

Figure~\ref{fig: tree-wc} shows that empirical ratios decrease with $ m $ and increase with $ M $ in tree networks when faced with hard instances for $ \gamma \in \{0.5, 2, 4\} $, which confirms our theoretical guarantees.
Specifically, we observe that when $ \gamma = 4 $, it delivers the best performance in both cases. Intuitively, a conservative strategy is beneficial when faced with a particularly challenging instance. Lastly, the empirical ratio is logarithmic in $ M $ when $ \gamma = 4 $, further verifying previous theoretical results.

\subsubsection{Path Selection with Cost}
For the online path selection with cost, we have shown that if a solution to Eq.~\eqref{eq_BVP_sufficiency} with parameter $\alpha$ exists, there is a corresponding online algorithm that achieves $\alpha$-competitive. However, Eq.~\eqref{eq_BVP_sufficiency} is notoriously difficult to analyze as a non-autonomous differential equation with singular boundary conditions. Therefore, we resort to find the smallest $\alpha$ such that a solution exists numerically, and show its logarithmic growth w.r.t. $\bar{p}$ in Figure~\ref{fig:alpha_rho}, in which the link capacity is set to $40$.


\section{Conclusions} \label{sec: conclusions}
In this paper, we investigated the role of the network topology and path lengths in determining the performance of a posted-price mechanism in the online path selection problem. We established new results about the dependence of competitive ratio on the path length bounds, recovered existing results about the dependence on the maximum path length, and elucidated particularly the varied influence of the path length bounds across different networks, in specific, line and hierarchical tree networks. Moreover, we studied the impact of system-level costs on the algorithm design and established sufficient and necessary conditions to be competitive. At last, we conducted extensive empirical experiments, which not only confirms our theoretical discovery but also uncovers the subtler effects of network structure on algorithmic performance for stochastic scenarios. These findings offer valuable insights for future development of more adaptive online algorithms that are tailored to specific network characteristics like bounded path lengths, and we hope it paves the way for further research in the domain of online algorithms against other types of constrained adversaries and other system-level coupling effects.

\bibliographystyle{IEEEtran}
\bibliography{mybib}

\end{document}